\documentclass[pra, showpacs, twocolumn, floatfix]{revtex4}
\usepackage{graphicx}
\usepackage{epstopdf}
\usepackage{amsmath, amsfonts, amssymb, bm}
\begin{document}
\title{Cooling a two-level emitter in photonic crystal environments}
\author{Marcela \surname{Cerbu}}

\author{Mihai A. \surname{Macovei}}
\affiliation{Institute of Applied Physics, Academy of Sciences of Moldova, 
Academiei str. 5, MD-2028 Chi\c{s}in\u{a}u, Moldova}

\author{Gao-xiang \surname{Li}}
\affiliation{Department of Physics, Huazhong Normal University, Wuhan 430079, China}
\date{\today}
\begin{abstract}
We investigate the resonant cooling phenomena of a driven two-level radiator embedded in a photonic crystal structure. 
We find that cooling occurs even at laser-atom-frequency resonance. This happens due to the atomic dressed-states 
coupling with different strengths resulting in population redistribution and cooling. Furthermore, for off-resonant driving 
the two-level particle can be in either ground or excited bare-state, respectively. This may help in engineering of novel 
amplified optical devices as well as highly coherent light sources.
\end{abstract}
\maketitle

\section{Introduction}
The mechanical action of light in resonant interaction with atoms was investigated for instance in \cite{sten}. 
There, the laser cooling techniques were described in details. Various developments of those ideas were 
further implemented \cite{rew} (and references therein). Additionally, laser cooling effects of trapped ions in a 
standing-  or a running-wave were studied in \cite{cir1}, while laser cooling of these systems via a squeezed 
environmental reservoir was investigated in Ref.~\cite{cir2}. Laser cooling of a trapped atom can be achieved 
in a optical cavity as well \cite{rew,cir3}. Cavity-mediated laser cooling was further investigated in \cite{beigeC}. 
Dark-state laser cooling of a trapped ion using standing waves and cooling 
by heating in the quantum optics domain were recently discussed in \cite{dark} and \cite{heat}, respectively. The 
experimental demonstration of ground-state laser cooling with electromagnetically induced transparency  
(EIT) \cite{eit_theor} was performed in \cite{eit_cool} for single-ion systems and in \cite{win} for an ion chain. 
Furthermore, double-EIT ground-state laser cooling without blue-sideband heating was proposed in \cite{ek}
while sideband cooling of atoms with the help of an auxiliary transition in \cite{yi_gxl}, and Raman 
sideband cooling of a single atom in an optical tweezer in  Ref.~{\cite{luk}}, respectively. Also, laser 
cooling of solids was addressed in \cite{sam} and \cite{nem}, respectively.

Now, there is an increased interest to apply the above mentioned or related techniques to cool artificially created 
systems like nano-mechanical oscillators or quantum circuit systems \cite{nori}. In particular, schemes to ground-state cooling 
of mechanical resonators were proposed in \cite{nmec}. Ground state cooling of a nanomechanical resonator in the 
nonresolved regime via quantum interference was the topic of Ref.~\cite{xe}. Cooling a mechanical resonator 
via coupling to a tunable double quantum dot was investigated in \cite{cool_dot}, while ground-state cooling of a 
mechanical resonator coupled to two coupled quantum dots was proposed in \cite{gxl}. The role of qubit dephasing in cooling 
a mechanical resonator by quantum back-action was discussed in \cite{back}. There it was shown that ground-state cooling 
of a mechanical resonator can only be realized if the superconducting flux qubit dephasing rate is sufficiently low.
A scheme for the cooling of a mechanical resonator via projective measurements on an auxiliary flux qubit which interacts with it 
was proposed in \cite{fast}. Furthermore, a flux qubit was experimentally cooled \cite{flq} using techniques somewhat related to 
the well-known optical sideband cooling methods. Finally, cooling a quantum circuit via coupling to a multiqubit ensemble was 
discussed in \cite{mma}.

Here, we investigate the cooling dynamics of a two-level atom via a photonic crystal environment. The two-level emitter 
is trapped in such a medium, while the linear dimensions of the trapping area inside the material have to be of the order 
of several emission wavelengths or even more. The trapping geometry has to be carefully engineered in order to the Born-Markov 
approximation being valid. The trapped two-level emitter is pumped with a moderately intense coherent  laser field that 
does not destroy the optical material due to insufficient applied intensities. When the corresponding generalized Rabi frequency 
is larger than the spontaneous decay rate, we have found an efficient cooling mechanism applicable even for resonant laser-atom 
interactions. The reason is various couplings of the involved atomic dressed-state transitions with the environmental reservoir leading 
to different decay rates that are also responsible for improving the cooling efficiency. While the two-level radiator is always in its lower 
dressed-state during cooling processes, in the bare-state frame it can be even in the excited state, i.e. the two-level atomic system is 
inverted. This situation is due to atom's coupling with the photonic crystal material and does not occur in free-space. On the other side, 
in free space or modified reservoirs efficient cooling occurs when the generalized Rabi frequency is additionally of the order of the vibrational mode 
frequency. Note that quantum dynamics in photonic crystal environments was/is widely investigated (see, for instance, 
\cite{phkr1,phkr2,phkr22,phkr23,phkr3} and references therein) including various possibilities for trapping of atoms in such media \cite{lukin,kimble}. 
Moreover, laser cooling of a trapped particle in a harmonic potential with increased Rabi frequencies was investigated as well in \cite{beige}. 

The article is organized as follows. In Sec. II we describe the analytical approach and the system of interest. We apply here standard 
procedures, generalized to photonic crystals environments, to arrive at a master equation describing the atomic vibrational degrees of 
freedom only. Sec. III deals with the corresponding equations of motion and discussion of the obtained results. Both the steady-state 
and time-dependent quantum dynamics are analyzed there as well. Finally, the Summary is given in the last section, i.e. Sec. IV.

\section{Analytical approach}
The Hamiltonian describing a two-level atomic system embedded in a photonic crystal environment and possessing an induced dipole $d$, frequency $\omega_{0}$,
and interacting with a coherent source of frequency $\omega_{L}$, in a frame rotating at $\omega_{L}$, and in the dipole approximation is \cite{kmek}:
\begin{eqnarray}
H &=& \sum_{k}\hbar\delta_{k} a^{\dagger}_{k}a_{k} +\hbar\Delta S_{z} +\hbar \Omega(S^{+}e^{i\vec k_{L}\vec r} + S^{-}e^{-i\vec k_{L} \vec r}) \nonumber \\
&+& i\sum_{k}(\vec g_{k}\cdot \vec d)(a^{\dagger}_{k}S^{-}e^{-i\vec k \vec r} - a_{k}S^{+}e^{i\vec k \vec r} ). \label{HM}
\end{eqnarray}
Here in the Hamiltonian (\ref{HM}), the first and the second terms describe, respectively,  the free environmental electromagnetic field (EMF) 
vacuum modes with $\delta_{k}=\omega_{k}-\omega_{L}$ as well as the free atomic energy with $\Delta=\omega_{0}-\omega_{L}$.
The third one characterizes the interaction of the two-level emitter localized at position $\vec r$ with the external coherent laser field, while
$\Omega$ being the corresponding Rabi frequency. The last term considers the qubit's interaction with environmental EMF
vacuum modes reservoir with $g_{k}$ describing the corresponding interaction strength. The atomic bare-state operators 
$S^{+}=|2\rangle \langle 1|$ and $S^{-}=[S^{+}]^{+}$ obey the commutation 
relations for su(2) algebra:
$[S^{+},S^{-}]=2S_{z}$ and $[S_{z},S^{\pm}]=\pm S^{\pm}$. Here, $S_{z}=(|2\rangle \langle 2| -
|1\rangle \langle 1|)/2$ is the bare-state inversion operator. $|2\rangle$ and $|1\rangle$ are the excited and
ground state of the qubit, respectively, while $a^{\dagger}_{k}$ and $a_{k}$ are the creation and the annihilation
operators of the EMF, and satisfy the standard bosonic commutation relations, i.e., $[a_{k},a^{\dagger}_{k'}]=\delta_{kk'}$, and 
$[a_{k},a_{k'}]=[a^{\dagger}_{k},a^{\dagger}_{k'}]=0$. 

We are interested in the laser dominated regime and transfer our investigations in the laser-dressed picture: 
$|2\rangle =\cos{\theta}|\bar 2\rangle - \sin{\theta}|\bar 1\rangle$ and $|1\rangle =\sin{\theta}|\bar 2\rangle + 
\cos{\theta}|\bar 1\rangle$ with $\cot{2\theta}=\Delta/(2\Omega)$. Further, the quantized vibrations of the atom's center-of-mass motion 
are represented as usual in the Lamb-Dicke limit, i.e. $k_{L}r=\eta(b+ b^{\dagger})$, where $b^{\dagger}(b)$ satisfy the single-mode bosonic 
commutation relations \cite{cir1,cir2,cir3}. Expanding the laser-atom interaction Hamiltonian to the second order in the small parameter $\eta$ as 
well as eliminating the EMF operators in the standard way and in the Born-Markov approximations \cite{kmek} one arrive then at the following 
dressed-state master equation:
\begin{eqnarray}
\frac{d}{dt}\rho &=& -i[H_{0},\rho] -\frac{\gamma_{0}}{4}\sin^{2}{2\theta}\{R^{2}_{z}\rho - 2R_{z}\bar\rho R_{z}+ \rho R^{2}_{z}\} \nonumber \\
&-&\gamma_{+}\cos^{4}{\theta}\{R^{+}R^{-}\rho - 2R^{-}\bar \rho R^{+} + \rho R^{+}R^{-} \} \nonumber \\
&-&\gamma_{-}\sin^{4}{\theta}\{R^{-}R^{+}\rho - 2R^{+}\bar \rho R^{-} + \rho R^{-}R^{+} \}, \label{MeqD}
\end{eqnarray}
where we have set $\hbar =1$ and
\begin{eqnarray}
&{}&H_{0}= \nu b^{\dagger}b +  \bar \Omega R_{z} + i\eta\Omega(R^{+} - R^{-})(b^{\dagger}+ b). \label{Hm0}
\end{eqnarray}
Here, the vibrational mode frequency is given by $\nu$ with $\nu b^{\dagger}b$ being its free energy. An additional second term, proportional to $\eta^{2}$, 
in Eq.~(\ref{Hm0}) leads to a shift of the oscillator frequency as well as to contributions proportional to higher orders in $\eta$ and, thus, was not taken into account here.
Further, the atomic dressed-basis operators are defined as: 
$R_{z}=|\bar 2\rangle \langle \bar 2| - |\bar 1\rangle \langle \bar 1|$, $R^{+}=|\bar 2\rangle \langle \bar1|$ and 
$R^{-}=|\bar 1\rangle \langle \bar 2|$ while obeying the commutation relations $[R^{+},R^{-}]=R_{z}$, and $[R_{z},R^{\pm}]=\pm 2R^{\pm}$, 
respectively. $\gamma_{\pm}$ and $\gamma_{0}$ are the single-atom dressed-state spontaneous decay rates at the frequencies 
$\omega_{L} \pm 2\bar \Omega$ and $\omega_{L}$ \cite{xxl}, respectively, while \cite{sten}
\begin{eqnarray}
\bar \rho = \frac{1}{2}\int^{1}_{-1}dx w(x)e^{i\eta(b+b^{\dagger})x}\rho e^{-i\eta(b+b^{\dagger})x}.
\label{brho}
\end{eqnarray}
The angular distribution of the spontaneous emission is characterized by the function: $w(x)=\frac{3}{4}(1+x^{2})$.
Notice that in free space environments $\gamma_{0}=\gamma_{\pm}$. This is not more the case if the atom 
is surrounded by a modified environmental electromagnetic field reservoir such as photonic crystals. Depending on 
density of modes distributions around the relevant dressed-state transitions these decay rates can differ 
substantially leading to inversion in the bare-states \cite{phkr2}. Finally, in Eq.~(\ref{MeqD}), one have 
performed the secular approximation, i.e. we have neglected the terms oscillating at the generalized Rabi 
frequency $\bar \Omega=\sqrt{\Omega^{2} + (\Delta/2)^{2}}$ or higher frequencies. This approximation 
is valid for $\bar \Omega \gg \gamma_{0,\pm}$, i.e. the spectral bands of the Mollow spectrum \cite{moll} are 
well distinguished.

In what follows, we shall apply the standard procedure to eliminate the atomic degrees of freedom from the Eq.~(\ref{MeqD}). 
This is valid when the atomic subsystem is faster than the vibrational ones. We proceed as follows: i) expand the exponents in Eq.~(\ref{brho})
to the second order in $\eta$; ii) go in a rotating frame at the generalized Rabi frequency $\bar \Omega$; iii) take the trace over atomic degrees of freedom.
Then one obtain the following master equation for the vibrational mode that still contains the atomic operators:
\begin{eqnarray}
\frac{d}{dt}\rho &=& \eta\Omega\bigl([B(t),\rho_{12}]e^{2i\bar \Omega t} -  [B(t),\rho_{21}]e^{-2i\bar \Omega t}\bigr) \nonumber \\
&-& \alpha\eta^{2}\bigl \{ (\gamma_{+}\cos^{4}{\theta}+\frac{\gamma_{0}}{4}\sin^{2}{2\theta})[B(t),B(t)\rho_{22}] \nonumber \\
&+&  (\gamma_{-}\sin^{4}{\theta}+\frac{\gamma_{0}}{4}\sin^{2}{2\theta})[B(t),B(t)\rho_{11}] + H.c. \bigr \}. \nonumber \\
\label{MeqT}
\end{eqnarray}
Here $\alpha=2/5$ and $B(t)=be^{-i\nu t} + b^{\dagger}e^{i\nu t}$ while $\rho_{\alpha\beta}=\langle \alpha|\rho_{a}|\beta\rangle \rho$ with $\{\alpha,\beta \in 1,2 \}$, 
and $\rho_{a}$ being the atomic density matrix operator only. To first order in $\eta$, from Eq.~(\ref{MeqD}), one can obtain:
\begin{eqnarray}
\rho_{21}(t)=\eta\Omega\{\bar B(t)\rho_{11}-\rho_{22}\bar B(t)\}e^{2i\bar \Omega t}, \label{rho21}
\end{eqnarray}
where $\rho_{12}(t)=(\rho_{21}(t))^{\dagger}$, and
$$
\bar B(t)=\frac{be^{-i\nu t}}{\Gamma_{\perp} + i(2\bar \Omega - \nu)} + \frac{b^{\dagger}e^{i\nu t}}{\Gamma_{\perp} + i(2\bar \Omega + \nu)},
$$
with $\Gamma_{\perp} = \gamma_{0}\sin^{2}{2\theta} + \gamma_{+}\cos^{4}{\theta} + \gamma_{-}\sin^{4}{\theta}$. Inserting Eq.~(\ref{rho21}) in Eq.~(\ref{MeqT})
and keeping only resonant contributions one arrive at the following master equation describing the vibrational mode alone:
\begin{eqnarray}
\frac{d}{dt}\rho = -A^{\ast}_{-}[b^{\dagger},b\rho] - A^{\ast}_{+}[b,b^{\dagger}\rho] + H.c., \label{meqv}
\end{eqnarray}
where $"\ast"$ means complex conjugation and
$$
A^{\ast}_{-}=\Gamma_{0} + \frac{(\eta\Omega)^{2}\langle R_{11}\rangle_{s}}{\Gamma_{\perp} + i(2\bar \Omega - \nu)}, ~
A^{\ast}_{+}=\Gamma_{0} + \frac{(\eta\Omega)^{2}\langle R_{22}\rangle_{s}}{\Gamma_{\perp} - i(2\bar \Omega - \nu)},
$$
with $\Gamma_{0}=\alpha\eta^{2}\bigl (\gamma_{-}\sin^{4}{\theta}\langle R_{11}\rangle_{s} + \gamma_{+}\cos^{4}{\theta}\langle R_{22}\rangle_{s} 
+ \frac{\gamma_{0}}{4}\sin^{2}{2\theta} \bigr )$. The steady-state values of the atomic operators $\langle R_{\alpha\alpha}\rangle_{s}=
\langle \alpha|\rho_{a}|\alpha \rangle_{s}$, $\alpha=\{1,2\}$, are obtained from Eq.~(\ref{MeqD}) without taking into account the vibrational degrees of 
freedom, namely \cite{xxl}:
\begin{eqnarray}
\langle R_{11}\rangle_{s} &=& \frac{\gamma_{+}\cos^{4}{\theta}}{\gamma_{+}\cos^{4}{\theta}+\gamma_{-}\sin^{4}{\theta}}, \nonumber \\
\langle R_{22}\rangle_{s} &=& 1 - \langle R_{11}\rangle_{s}. \label{s_s} 
\end{eqnarray}
That is, we have considered that the vibrational degrees of freedom do not modify the population distribution among the involved dressed-states that are governed 
by the external applied coherent laser field.
\begin{figure}[t]
\centering
\includegraphics[height=3cm,width=8.5cm]{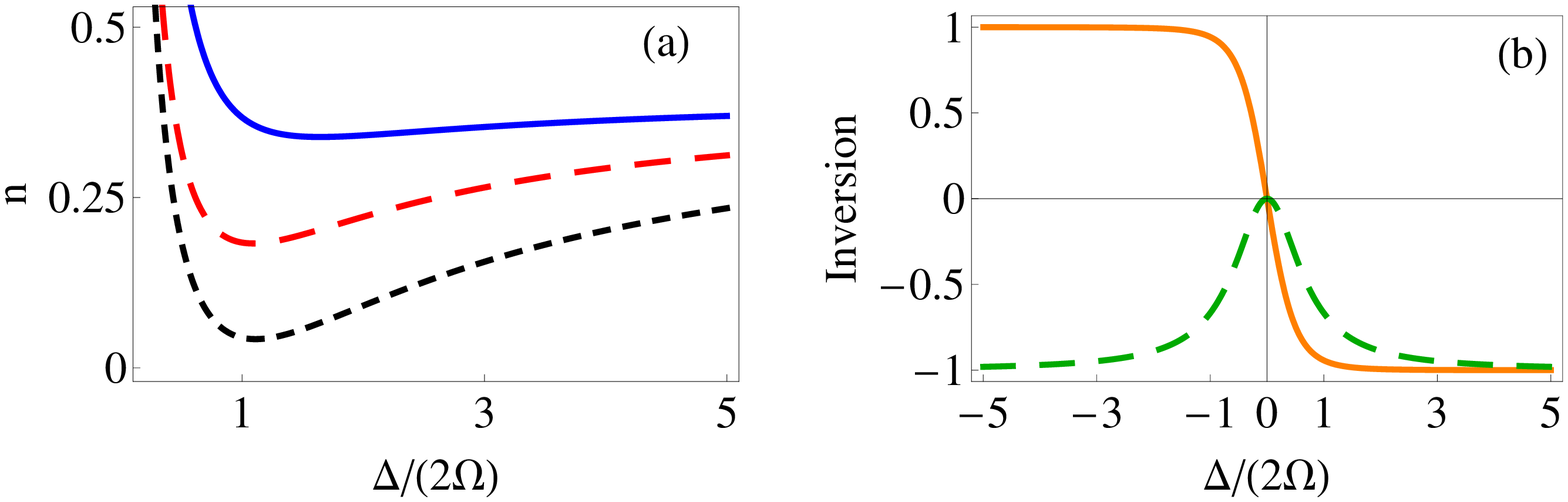}
\caption{\label{fig-1} 
(color online) The steady-states of (a) the mean-phonon number $n=\langle b^{\dagger}b\rangle_{s}$ and (b) dressed-state $\langle R_{z}\rangle_{s}$ (solid line) 
as well as bare-state $2\langle S_{z}\rangle_{s}$ (long-dashed curve) inversion operator versus normalized detuning. Here, $\eta=0.1$ whereas 
$\Omega/\gamma = 5$ with $\gamma_{+}=\gamma_{-}=\gamma_{0}\equiv \gamma$. In (a) the solid, long-dashed and short-dashed lines are for 
$\nu/\gamma=2$, $6$ and $12$, respectively.}
\end{figure}

In the next section, with the help of Eq.~(\ref{meqv}), we shall describe the cooling processes of a pumped two-level emitter embedded in a photonic crystal.
\section{Cooling phenomenon}
In the Heisenberg picture, the master equation~(\ref{meqv}) can be represented as follows:
\begin{eqnarray}
\frac{d}{dt}\langle Q\rangle = -A^{\ast}_{-}\langle[Q,b^{\dagger}]b\rangle - A^{\ast}_{+}\langle[Q,b]b^{\dagger}\rangle +H.c.,
\label{MQ}
\end{eqnarray}
where $Q$ is any vibrational operator. Note that, in general, for the non-Hermitian operators $Q$, the H.c. terms should be evaluated without
conjugating $Q$, i.e., by replacing $Q^{\dagger}$ with $Q$ in the Hermitian conjugate parts. With the help of Eq.~(\ref{MQ}), one can easily 
obtain the quantum dynamics of the mean phonon number in the vibrational mode, that is:
\begin{eqnarray}
\frac{d}{dt}\langle b^{\dagger}b\rangle = -\bigl(A^{(-)}-A^{(+)} \bigr)\langle b^{\dagger}b\rangle + A^{(+)}, 
\label{bm}
\end{eqnarray}
where $A^{(-)}=A_{-} + A^{\ast}_{-}$, while $A^{(+)}=A_{+} + A^{\ast}_{+}$. Evidently, cooling occurs for $A^{(-)} > A^{(+)}$. Particularly, the
cooling rate $C = A^{(-)} - A^{(+)}$ is:
\begin{eqnarray}
C= - \frac{2(\eta\Omega)^{2}\Gamma_{\perp}\langle R_{z}\rangle_{s}}{\Gamma^{2}_{\perp} + (2\bar \Omega - \nu)^{2}},
\label{coolC}
\end{eqnarray}
where the steady-state dressed-state inversion operator is given by $\langle R_{z}\rangle_{s} =\langle R_{22}\rangle_{s} - \langle R_{11}\rangle_{s}$. 
One can observe that cooling occurs always when the two-level emitter is in the lower dressed-state, i.e. $\langle R_{z}\rangle <0$. 
In the steady-state, we have from Eq.~(\ref{bm}) and Eq.~(\ref{coolC}) that $\langle b^{\dagger}b\rangle_{s} = A^{(+)}/C$ or: 
\begin{eqnarray}
\langle b^{\dagger}b\rangle_{s} &=& \frac{\langle R_{22}\rangle_{s}}{\langle R_{11}\rangle_{s}-\langle R_{22}\rangle_{s}} \nonumber \\
&+& \frac{\Gamma_{0}\bigl( \Gamma^{2}_{\perp} + (2\bar \Omega-\nu)^{2}\bigr )}
{(\eta\Omega)^{2}\Gamma_{\perp}\bigl( \langle R_{11}\rangle_{s} - \langle R_{22}\rangle_{s} \bigr)}.
\label{bpm}
\end{eqnarray}
\begin{figure}[t]
\centering
\includegraphics[height=3cm,width=8.5cm]{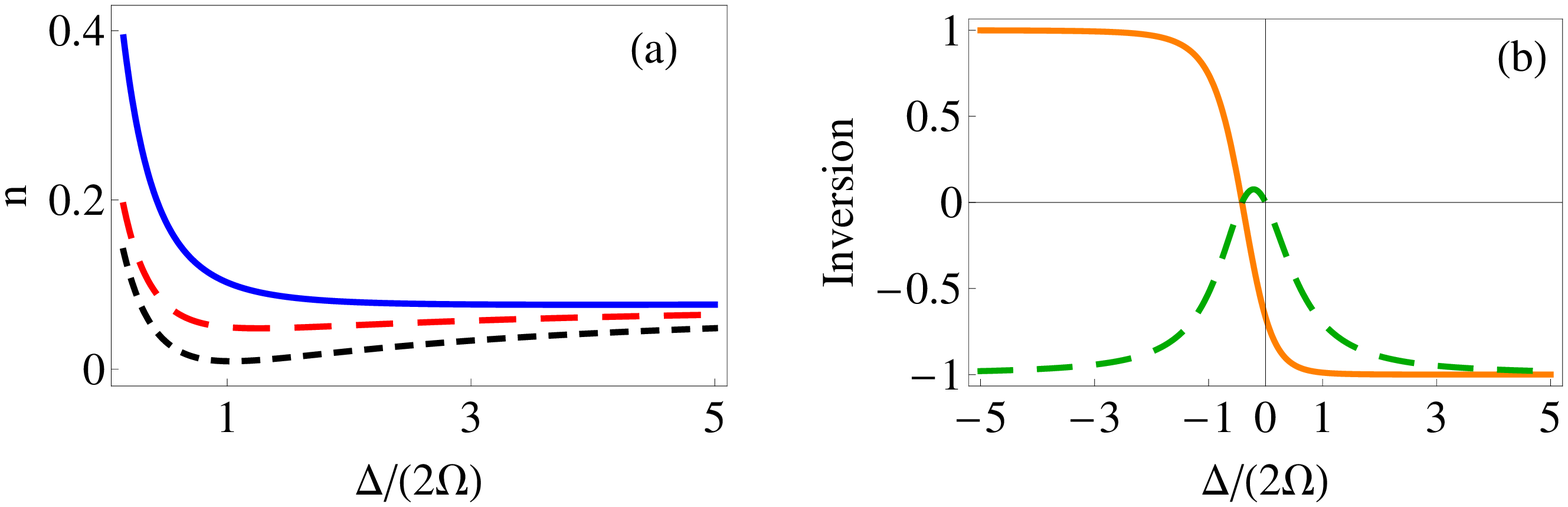}
\caption{\label{fig-1e} 
(color online) Same as in Fig.~(\ref{fig-1}) with $\Omega/\gamma_{+}=5$, $\gamma_{-}/\gamma_{+}=\gamma_{0}/\gamma_{+}=0.2$.
In (a) the solid, long-dashed and short-dashed curves are for $\nu/\gamma_{+} = 2$, $6$ and $12$, respectively.}
\end{figure}

In the next subsections, we shall rigorously analyze the steady-state as well as the time-dependent quantum dynamics of the cooling 
effect characterizing the particular system described here. 

\subsection{The steady-state cooling dynamics}
Fig.~(\ref{fig-1}a) shows the mean-number of vibrational quanta according to Eq.~(\ref{bpm}) and for standard situations, that is, when the pumped two-level
emitter interacts with the usual vacuum modes of the environmental electromagnetic field reservoir \cite{cir1,beige}. Cooling efficiency improves, i.e. $n \ll 1$, if the 
generalized Rabi frequency $2\bar \Omega$ approaches the vibrational frequency $\nu$ while the external pumping coherent field is evidently off-resonant. 
As it was already mentioned, cooling always occurs when the two-level atom is in its ground dressed-state, and for free space, the atom is also in its ground 
bare-state, respectively, (see Fig.~\ref{fig-1}b where the bare-state inversion $\langle S_{z}\rangle_{s}=\cos{2\theta}\langle R_{z}\rangle_{s}/2$ is 
shown as well). For the sake of comparison, Fig.~(\ref{fig-1e}a) shows respectively the cooling dynamics when the two-level emitter is surrounded by a photonic 
crystal environment. The cooling efficiency improves and it is visible there. The steady-state atomic population behaviors differ from the free space. Particularly, 
inversion in the bare-state frame occurs (see Fig.~\ref{fig-1e}b). Therefore, we shall further look in more details on the cooling phenomenon via photonic crystals 
environments.
\begin{figure}[t]
\centering
\includegraphics[height=3cm,width=8.5cm]{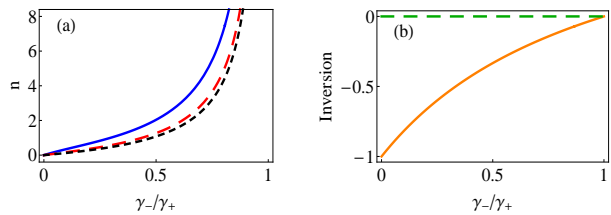}
\caption{\label{fig-2} 
(color online) The steady-states of (a) the mean-phonon number $n=\langle b^{\dagger}b\rangle_{s}$ and (b) dressed-state $\langle R_{z}\rangle_{s}$ (solid line) 
as well as bare-state $\langle S_{z}\rangle_{s}$ (long-dashed curve) inversion operator as a function of $\gamma_{-}/\gamma_{+}$. Here, 
$\Delta=0$ and $\gamma_{0} \equiv \gamma_{-}$ while all other parameters are the same as in Fig.~(\ref{fig-1}).}
\end{figure}
\begin{figure}[b]
\centering
\includegraphics[height=3cm,width=8.5cm]{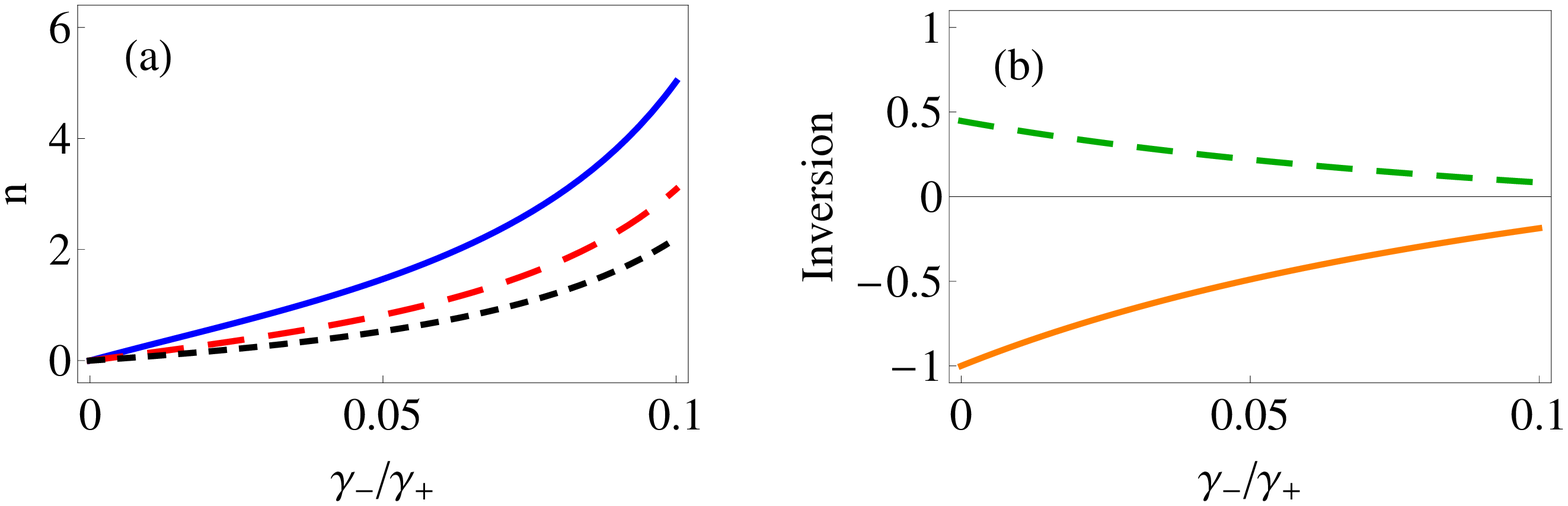}
\caption{\label{fig-3} 
(color online) The steady-states of (a) the mean-phonon number $n=\langle b^{\dagger}b\rangle_{s}$ and (b) dressed-state $\langle R_{z}\rangle_{s}$ (solid line) 
as well as bare-state 2$\langle S_{z}\rangle_{s}$ (long-dashed curve) inversion operator as a function of $\gamma_{-}/\gamma_{+}$. Here, 
$\Delta/(2\Omega)=-0.5$ while all other parameters are the same as in Fig.~(\ref{fig-2}).}
\end{figure}

In this respect, Fig.~(\ref{fig-2}a) depicts the mean-vibrational-phonon number when the two-level emitter is embedded in a modified environmental electromagnetic field 
reservoir such as photonic crystals. Interestingly, cooling occurs even at laser-atom resonance. This happens because the decay rates at various involved 
dressed-state transitions can differ substantially which is not the case in free space at resonance.  Here, again, the atom is in its lower dressed-state 
while the bare-state inversion is zero meaning that the atom is equally distributed on the two bare states (see Fig.~\ref{fig-2}b). Furthermore,  in Fig.~(\ref{fig-3}), we show 
the cooling processes when the two-level emitter can be even in the excited bare-state. This may be of particular interest
for engineering of highly coherent laser sources, for instance, where the vibrational degrees of freedom do not influence the photon statistics. Efficient 
cooling via modified environmental electromagnetic field reservoirs occurs also for positive detunings although the atom will be in its lower dressed/bare-state, 
respectively. Thus, cooling takes place for both positive or negative detunings. This feature is not proper for free space setups, i.e. when the two-level atom is 
surrounded by the usual vacuum modes of the environmental electromagnetic field reservoir. The explanation is as follows: the laser-dressed atom decays on 
transition $|\bar 2\rangle \to |\bar 1\rangle$ with a decay rate $\gamma_{+}\cos^{4}{\theta}$, while on $|\bar 1\rangle \to |\bar 2\rangle$ transition
with $\gamma_{-}\sin^{4}{\theta}$, respectively. Cooling occurs when the two-level emitter is in its lower dressed-state, $|\bar 1\rangle$, because in this 
state the phonon generation processes are minimized, and, therefore, $\gamma_{+}\cos^{4}{\theta}$ should be larger than $\gamma_{-}\sin^{4}{\theta}$. 
In free space $\gamma_{+}=\gamma_{-}$ and, thus, cooling is achieved for positive detunings only since in this case $\cos^{4}{\theta} > \sin^{4}{\theta}$. 
Note that $\cos^{2}{\theta}=(1+\Delta/(2\bar \Omega))/2$ while $\sin^{2}{\theta}=(1 - \Delta/(2\bar \Omega))/2$. However, due to coupling of the 
dressed-atom with photonic crystal  environments, cooling can take place for negative detunings as well and still 
$\gamma_{+}\cos^{4}{\theta} > \gamma_{-}\sin^{4}{\theta}$, i.e. when $\gamma_{+} \gg \gamma_{-}$.

\subsection{The time-dependent cooling dynamics}
The time-dependent quantum dynamics for the mean values of the dressed-state atomic inversion operator, dressed-state coherences, as well as the mean-phonon number 
of vibrational quanta is described by the following expressions (see Eq.~\ref{MeqD} and Eq.~\ref{bm}):
\begin{eqnarray}
\langle R_{z}(t)\rangle &=& \bigl(\langle R_{z}(0)\rangle - \langle R_{z}\rangle_{s}\bigr)e^{-2\gamma_{s}t} + \langle R_{z}\rangle_{s}, \nonumber \\
\langle R^{+}(t)\rangle &=&\langle R^{+}(0)\rangle e^{-\Gamma_{\perp}t},~ {\rm with}~  \langle R^{-}\rangle = \langle R^{+}\rangle^{\dagger}, \nonumber \\
\langle b^{\dagger}b\rangle_{t} &=& \bigl(\langle b^{\dagger}b\rangle_{0} - \langle b^{\dagger}b\rangle_{s}\bigr)e^{-Ct} + \langle b^{\dagger}b\rangle_{s},
\label{td_ex}
\end{eqnarray}
where $\gamma_{s}=\gamma_{+}\cos^{4}{\theta}+\gamma_{-}\sin^{4}{\theta}$, while $\langle R_{z}(0)\rangle=\cos{2\theta}\langle S_{z}(0)\rangle 
+\sin{2\theta}\bigl(\langle S^{+}(0)\rangle + \langle S^{-}(0)\rangle \bigr)$, $\langle R^{+}(0)\rangle=\cos^{2}{\theta}\langle S^{+}(0)\rangle - 
\sin^{2}{\theta}\langle S^{-}(0)\rangle - \sin{2\theta}\langle S_{z}(0)\rangle$ and $\langle b^{\dagger}b\rangle_{0}$ are the initial conditions for the mean values of 
atomic inversion operator, coherences, and vibrational phonon number, respectively. Notice that the expressions (\ref{td_ex}) are valid for $t \gg (2\bar \Omega)^{-1}$
and $\eta\Omega < {\rm Max}\{\gamma_{\pm},\gamma_{0}\} \ll 2\bar \Omega$. Furthermore, our cooling approach requires that $\{2\gamma_{s},\Gamma_{\perp} \} \gg C$ 
which, in principle, can always be arranged. Particularly, for $\{\Gamma_{\perp},(2\bar \Omega - \nu)\} \sim \gamma_{+}$ and $\langle R_{z}\rangle_{s} \approx -1$ one has
$C \approx (\eta \Omega/\gamma_{+})^{2}\gamma_{+}$. When $\Omega/\gamma_{+} =5$ and $\eta=0.1$, cooling occurs for $t \gg 4/\gamma_{+}$. 
For typical values of spontaneous decay rates at optical frequencies, $\gamma_{+} \sim 10^{8}-10^{9}{\rm Hz}$, cooling is achieved in microseconds.

\section{Summary}
In summary, we have investigated the cooling efficiency of a laser-pumped two-level 
emitter placed in a modified surrounding electromagnetic field reservoir like photonic 
crystals. Particularly, cooling occurs for positive or negative laser-atom detunings as 
well as at resonance. Furthermore, the two-state particle can be even in the 
excited bare-state during cooling processes facilitating sensitive applications towards 
better coherent light sources as well as light amplification. 

\acknowledgments
G.-x. Li is grateful to the financial support from National Natural Science Foundation of China (Grant No. 61275123) 
and the National Basic Research Program of China (Grant No. 2012CB921602).


\end{document}